\documentclass[aps,prb,twocolumn,superscriptaddress]{revtex4-1}
\usepackage{graphicx}
\graphicspath{ {Images/}}

\newcommand{\beq}{\begin{equation}}
\newcommand{\eeq}{\end{equation}}
\newcommand{\sgn}{\mathop{\mathrm{sgn}}}
\newcommand{\DO}{{{\rho}}}
\newcommand{\OpCalH}{\ensuremath{\hat {\mathcal{H}}}\xspace}
\newcommand{\TiH}{\ensuremath{ {\tilde{H}}}\xspace}
\newcommand{\Flux}{\ensuremath{\hat{\Phi}}}
\newcommand{\Charge}{\ensuremath{\hat{Q}}}

\usepackage{datetime}
\usepackage{xspace}
\usepackage{textcomp}
\usepackage{mathtools}
\usepackage{makeidx}
\usepackage{mathrsfs}
\usepackage{amsmath}

\newcommand{\COM}[2] {\ensuremath{\left[ {#1} , {#2}\right]}\xspace}
\newcommand{\ACOM}[2] {\ensuremath{\left\{ {#1} , {#2}\right\}}\xspace}

\newcommand{\Half} {\ensuremath{\frac{1}{2}}\xspace}

\newcommand{\EX}[1] {\ensuremath{\left\langle #1 \right\rangle}\xspace}

\newcommand{\ud}{\mathrm{d}} 
\newcommand{\ui}{\mathrm{i}} 

\DeclareMathOperator{\Tr}{Tr}
\newcommand{\Trace}[1] {\ensuremath{\Tr \left[ #1 \right]}\xspace}

\newcommand{\be}{\begin{equation}}
\newcommand{\bel}[1]{\begin{equation}\label{#1}}
\newcommand{\ee}{\end{equation}}
\newcommand{\ba}{\begin{eqnarray}}
\newcommand{\ea}{\end{eqnarray}}
\newcommand{\bal}{\begin{align}}
\newcommand{\eal}{\end{align}}

\newcommand{\TD}[2]{\ensuremath{\frac{\ud #1}{\ud #2}}\xspace}

\newcommand{\St}{$^\mathrm{st}$\xspace}
\newcommand{\Nd}{$^\mathrm{nd}$\xspace}

\newcommand{\Fig}[1]{Fig.~\ref{#1}\xspace}
\newcommand{\Eq}[1]{Eq.~(\ref{#1})\xspace}

\newcommand{\OpQ}{\ensuremath{\hat Q}\xspace}

\newcommand{\OpP}{\ensuremath{\hat P}\xspace}
\newcommand{\OpA}{\ensuremath{\hat A}\xspace}
\newcommand{\Opa}{\ensuremath{\hat a}\xspace}

\newcommand{\OpB}{\ensuremath{\hat B}\xspace}

\newcommand{\OpL}{\ensuremath{\hat {L}}\xspace}

\newcommand{\OpX}{\ensuremath{\hat X}\xspace}

\newcommand{\OpH}{\ensuremath{\hat H}\xspace}

\newcommand{\QSERG}{Quantum Systems Engineering Research Group, Loughborough University, Loughborough, Leicestershire LE11 3TU, United Kingdom}
\usepackage{amssymb,amsmath,mathrsfs,dsfont,color}
\newcommand{\LBORO}{Department of Physics, Loughborough University}
\begin{document}
 
\title{Some implications of superconducting quantum interference to the application of master equations in engineering quantum technologies}

\author{S.N.A.~Duffus}
\affiliation{\QSERG}
\affiliation{\LBORO}
\author{K.N.~Bjergstrom}
\affiliation{\QSERG}
\affiliation{\LBORO}
\author{V.M.~Dwyer}
\affiliation{\QSERG}
\affiliation{The Wolfson School, Loughborough University}
\author{J.H.~Samson}
\affiliation{\QSERG}
\affiliation{\LBORO}
\author{T.P.~Spiller}
\affiliation{York Centre for Quantum Technologies, Department of Physics, University of York, York, YO10 5DD, United Kingdom}
\author{A.M.~Zagoskin}
\affiliation{\LBORO}
\author{W.J.~Munro}
\affiliation{NTT Basic Research Laboratories, NTT Corporation, 3-1~Morinosato-Wakamiya, Atsugi, Kanagawa 243-0198, Japan}
\author{Kae Nemoto}
\affiliation{National Institute of Informatics, 2-1-2 Hitotsubashi, Chiyoda-ku, Tokyo 101-8430, Japan}
\author{M.J.~Everitt}
\email{m.j.everitt@lboro.ac.uk}
\affiliation{\QSERG}
\affiliation{\LBORO}

\date{\today}

\begin{abstract}
In this paper we consider the modelling and simulation of open quantum systems from a device engineering perspective. We derive master equations at different levels of approximation for a Superconducting Quantum Interference Device (SQUID) ring coupled to an ohmic bath. We demonstrate that the master equations we consider produce decoherences that are qualitatively and quantitativly dependent on both the level of approximation and the ring's external flux bias. We discuss the issues raised when seeking to obtain Lindbladian dissipation and show, in this case, that the external flux (which may be considered to be a control variable in some applications) is not confined to the Hamiltonian, as often assumed in quantum control, but also appears in the Lindblad terms. 
\end{abstract}


\maketitle

\section{Introduction}

With its ability to provide substantial cost savings and speed up the exploration of parameter space, modelling and simulation plays a central role in the engineering process. As Quantum Technologies (QTs) move away from laboratory demonstrations and become integrated into consumer systems, accurate modelling will become increasingly important \cite{Dowling2003,Boixo2014,Clarke2004,Ruggiero2003,Ogunyanda2015}. Here robust, and generally hierarchical, quantitative simulations will be required which are capable of accurately and reliably predicting the behaviour of the system-under-development at different levels of abstraction. The ultimate ambition of this approach being to achieve a level of realism that would enable the sort of zero-prototyping that occurs in the design of Very Large Scale Integrated (VLSI) microelectronics and which is also now becoming an aim of the automotive and other industries. Given the intractability by classical means of modelling complex quantum systems, it is an open question as to how well and how far this design paradigm can be translated to the engineering of quantum technologies. Consequently, there is a need to investigate the extent to which it is possible to develop a hierarchy of system models that is able to provide, from a design perspective, usefully accurate modelling, simulation and figures of merit at the component, device and system level. 

Before one might consider developing such a system level view, it is also necessary to establish the effectiveness of existing device level models and the degree to which these might be leveraged for such applications. Of particular interest, at this stage, is the quantitative accuracy of models of open systems for single quantum objects, such as the case of a classical device acting as the environment for some quantum component. Ultimately such models will need to include time-varying parameters such as in the case, for example, of the feedback and control of a quantum resource.
One standard approach, that might prove effective in forming part of an engineering design strategy, derives from the application of quantum master equations, as these provide a generic pathway for the modelling of a quantum system and its interaction with the environment. Master equations have become a standard tool in this regard as they promise a means of extracting system properties from environmental influences. It is a general view that the dynamics described is in good qualitative agreement with the ensemble average of the system being studied, and that deviations of theory from experimental observations can be brought into acceptable line by fine tuning model parameters, leading to the conclusion that master equations provide a good phenomenological approach \cite{Weiss1999,Joos2003,Santos2014}. The most widely used master equations are memoryless, and take the Lindblad form \cite{Lindblad1976,Gardiner1991,Breuer2002, Schloss2007}
\beq \label{eq:meqLindblad}
\frac{d\DO_S}{dt} = -\frac{\ui}{\hbar}[\OpH_S,\DO_S] + \frac{1}{2}\sum_{j}\bigg\{[\OpL_{j}, \DO_S \OpL_{j}^{\dagger}] + [\OpL_{j}\DO_S, \OpL_{j}^{\dagger}]\bigg\}
\eeq
where $\DO_S$ is the reduced density operator of the system, $\OpH_S$ is the system Hamiltonian and the $\OpL_{j}$ account for the effects of the environmental degrees of freedom. Lindblad master equations dominate work on open quantum systems as they conserve probability (i.e. $\Trace{\rho_S}=1$) and ensure that $\DO_S$ is at least physically acceptable (i.e. there are no negative probabilities, etc.). Master equations of non-Lindblad form, on the other hand, usually will lead to some situations which are unphysical \cite{Gardiner1991,Breuer2002, Schloss2007, Munro1996, Gao1997} 

In this work we seek to explore how effective the master equation approach might be in engineering superconducting quantum devices, and in particular for the case of the Superconducting Quantum Interference Device (SQUID) ring (an $LC$ circuit enclosing a Josephson junction weak link) coupled to a low temperature Ohmic bath, with cut-off frequency $\Omega$. We note that Josephson junction based devices are currently of significant technological importance, with applications in quantum computation (e.g. D-Wave, IBM and Google) and metrology. Beyond their significance for emerging quantum technologies, there are two further reasons we have chosen to investigate the decoherence of SQUIDs as an example Josepheson junction device.

First, the contribution to the Hamiltonian of the Josephson junction term brings with it non-trivial mathematical properties which test the suitability of master equations to quantitative engineering applications (including potentially control through the externally applied flux $\Phi_x$). Recent work has provided an exact solution to the similar (but simpler) Quantum Brownian Motion (QBM) problem (in a quadratic well) to all orders of Born Approximation. The solution \cite{Fleming2011, Massignan2015} displays a logarithmic dependence on $\Omega$ which indicates the general result for such problems that the limit $\Omega \rightarrow \infty$ does not exist (i.e. $\Omega$ is finite) and, additionally, highlights the importance of parameterising the bath properly. The common practice of terminating master equations at first order in $\omega_0/\Omega$ (where $\omega_0=1/\sqrt{LC}$ is a characteristic frequency in the system) assumes that an expansion to second order will only produce small corrections. 

The second reason for our choice of system is that it allows us to investigate the issues in the standard derivation of the master equation for a SQUID/Ohmic environment for a hierarchy of models, in which $\omega_0/\Omega$ plays the role of a small expansion parameter. Thus, first and second order master equations are obtained, using what might be termed standard techniques, and compared through quantities at the steady state, such as purity and screening current. The difficulties in such analysis are discussed and the generally bespoke nature of such methods highlighted. Finally, while a higher order Born series approximation might be more valuable, the issues which arise in the current, simpler analysis are quite significant enough and are likely to be indicative of those considerations that an investigation of stronger coupling through a Born series would require.  
     
\section{Model - A SQUID with a Lossy Bath}

The system considered here consists of a SQUID ring coupled to an Ohmic bath represented by an infinite number of harmonic oscillators at absolute zero temperature. Ideally, the Hamiltonian for this system should be derived from a full quantum field theoretic description or from a general quantum circuit model (see, for example,~[\onlinecite{Burkard2005}]), and such analysis would certainly be needed for any application of this method to the engineering of a specific quantum device, however its inclusion here would complicate our presentation and distract from our central discussion of the issues associated with deriving master equations for superconducting systems. The Hamiltonian for the system is therefore taken to be of the form, $\OpCalH = \OpCalH_{S} +\OpCalH_{B}+\OpCalH_{I} $, which is simply the sum of the Hamiltonians of the SQUID $\OpCalH_{S}$, the bath $\OpCalH_{B}$ and the interaction between them $\OpCalH_{I}$, given by:
\beq\begin{split}\label{eq:1}
	\OpCalH_{S}&=\frac{\hat{Q}^{2}}{2C}+\frac{\left(\hat{\Phi}-\Phi_{x}\right)^{2}}{2L}-\hbar\nu\cos{\left(\frac{2\pi\hat{\Phi}}{\Phi_0}\right)} \\ 
	\OpCalH_{B}&= \sum_{n}\frac{\hat{Q}^{2}_{n}}{2C_{n}}+\frac{\hat{\Phi}_n^{2}}{2 L_n}\\ 
	\OpCalH_{I}&= 	-\left(\hat{\Phi}-\Phi_{x}\right)\sum_{n}\kappa_{n}\hat{\Phi}_{n} \end{split}\eeq
where \Charge, $\Charge_{n}$, $\Flux$ and $\Flux_{n}$ ($n = 1, 2, ...$) represent the charge and flux operators of the system and bath modes respectively, so that $\COM{\Flux}{\Charge}=\COM{\Flux_n}{\Charge_n}=i\hbar$, $\Phi_x$ is an externally applied flux, and $L, C$ and the $L_n,C_{n}$ are the inductance and capacitance values in each subsystem. As the Hamiltonian has not been derived from a complete circuit model, the parameters must be considered as being the effective values that arise through the coupling of the components together - thus for example $L$ and the $L_{n}$ are effective inductances.   
 The bath mode coupling strength $\kappa_{n}$ is related to a system damping rate $\gamma$ through the explicit expression of the bath spectral density and correlation functions \cite{Breuer2002}. We note that, as is usually the case with this sort of `particle confined by a potential' system, we have not included any capacitive (momentum) coupling; its inclusion would naturally change the analysis which follows. 

The SQUID Hamiltonian may be simplified to that of an unshifted harmonic oscillator plus a perturbation term through the unitary translation operator $\hat{T}= \exp{\left(-\ui \hat{Q}\Phi_{x}/\hbar\right)}$. The system Hamiltonians acting in the translated (external flux) basis may then be written as \cite{Tannoudji2005,Everitt2001,Everitt2001a}:
 \ba  
 \OpH'_{S}&=& \hat{T}^{\dagger}\OpCalH_{S}\hat{T} =\frac{\hat{Q}^{2}}{2C}+\frac{\hat{\Phi}^{2}}{2L}-\hbar\nu\cos{\left(\frac{2\pi}{\Phi_0}\left(\hat{\Phi}+\Phi_{x}\right)\right)}
\nonumber  
\\ \nonumber
	\OpH'_{B}&=&   \hat{T}^{\dagger}\OpCalH_{B}\hat{T}=\OpH_{B}=\sum_{n}\frac{\hat{Q}^{2}_{n}}{2C_{n}}+\frac{\hat{\Phi}_n^{2}}{2 L_n} 
\\ \label{eq:HamFluxBasis}
 \OpH'_{I}&=&    \hat{T}^{\dagger}\OpCalH_{I}\hat{T}=-\Flux\sum_{n}\kappa_{n}\hat{\Phi}_{n} = -\Flux \OpB	\ea
where we have introduced $\OpB$ as a shorthand for the bath operator $\sum_{n}\kappa_n \hat{\Phi}_n$ and will drop the primed notation from now on. As usual, as long as there is no explicit time dependence in the total Hamiltonian $\OpH = \OpH_{S} +\OpH_{B}+\OpH_{I}$, the Schr\"odinger and the Liouville-von Neumann equations are unaltered by the translation. If the external flux is time dependent there will arise additional terms in $\OpH'_{S}$ due to this translation of the form $\OpQ \dot \Phi_x$ - however these would be small in the adiabatic limit\cite{Clark1998}.

 \section{Review of Deriving The General Form of The Master Equation}
The derivation of the master equation can now follow standard textbook methods, we include this discussion for coherence within the paper, however the reader who is familiar with such material may wish to move forward to section~\ref{IME}. The dynamics of the system+bath is given by the  Liouville-von Neumann equation:
\beq \label{eq:LVN}
	\TD{\DO(t)}{t}=-\frac{\ui}{\hbar}[\OpH,\DO(t)]
\eeq
As it is not generally possible to solve this equation, analytically or numerically, we derive a master equation that approximates the dynamics of the reduced density matrix $\tilde\DO_S(t)$ for the SQUID ring. Rotating the system into the interaction picture, \Eq{eq:LVN} becomes:
\beq \label{eq:LVNI}
	\TD{\tilde{\DO}(t)}{t}=-\frac{\ui}{\hbar}\COM{\TiH_I(t)}{\tilde{\DO}(t)}
\eeq
where we define $\tilde{A}=e^{{\ui (\OpH_{S}+\OpH_B)t}/{\hbar}}\OpA e^{{-\ui (\OpH_{S}+\OpH_B)t}/{\hbar}}$ as the rotated version of an operator \OpA. Integrating ~\Eq{eq:LVNI} yields:
\bel{eq:IntegratedLVNI}
\tilde{\DO}(t) = \tilde{\DO}(0)-\frac{\ui}{\hbar}\int_{0}^t \ud s[\tilde{H}_{I}(s),\tilde{\DO}(s)]
\ee
It is usual, at this stage, to apply a set of assumptions which are collectively known as the Born-Markov approximation. This starts with the assumption that, at some time in the past which we label $t=0$, the bath and system were uncorrelated, i.e. in a separable pure state, so that $\tilde{\DO}(0)=\tilde{\DO}_{S}(0)\otimes\tilde{\DO}_{B}(0)$ where $\tilde{\DO}_{S}$ and $\tilde{\DO}_{B}$ are the reduced density matrices for the SQUID ring and bath respectively. This approximation is generally sound in quantum optics but may not hold so well for condensed matter systems. It is not clear whether non-Markovian master equations will become necessary in such cases, however these bring with them a number of additional challenges that are beyond the scope of this work. For now we impose the \textit{uncorrelated} assumption and we justify it as being valid at the point that the superconducting condensate first forms. That is, if the condensation process removes any existing correlations between the electrons and their environment, then this approximation is acceptable and $t=0$ is taken to be the time at condensation.

Substituting the expression for $\tilde{\rho}(t)$ into the Liouville-von Neumann equation in the interaction picture,~\Eq{eq:LVNI} gives:
\beq
\label{eq:4a}
	\TD{\tilde{\rho}(t)}{t}=-\frac{\ui}{\hbar}[\tilde{H}_{I}(t),\tilde{\rho}(0)]-\frac{1}{\hbar^{2}}\int_0^t \ud s[\tilde{H}_{I}(t),[\tilde{H}_{I}(s),\tilde{\rho}(s)]]
\ee
If further we apply the standard Markovian restriction that the bath is memoryless, it is possible to extend this to $\tilde{\DO}(t)=\tilde{\DO}_{S}(t)\otimes\tilde{\DO}_{B}(t)$, although previous studies of fully quantum mechanical models of electromagnetic fields with SQUID rings show that there may be significant back-action between the ring and its environment which cannot be captured by this approximation\cite{Helmer2009,Romero2009,Stiffell2005,Everitt2001,Everitt2001a} . However, it does allow for a further assumption that the bath is sufficiently big that the SQUID ring will have a negligible effect on it, so that we may take $\tilde{\DO}_{B}(t)$ as approximately constant. 

Such considerations already raise the prospect that the Born-Markov approximation may be inadequate for the accurate study of condensed matter systems, limiting the use of master equations in the modelling and simulation for quantitive applications as part of an engineering solution; at best they may offer only a phenomenological tool. Despite these difficulties, such phenomenological models are important and an investigation of their predictions is still worthwhile and we proceed on that basis. The consequence is that \Eq{eq:4a} simplifies to:
\ba 
	\TD{\tilde{\DO}_S (t)}{t}\otimes\tilde{\DO}_{B}&=&-\frac{\ui}{\hbar}[\tilde{H}_{I}(t),\tilde{\DO}_{S}(0)\otimes \tilde{\DO}_{B}]  \\ \nonumber   &&
	-\frac{1}{\hbar^{2}}\int_0^t \ud s[\tilde{H}_{I}(t),[\tilde{H}_{I}(s),\tilde{\DO}_{S}(s)\otimes\tilde{\DO}_{B}]]
\ea
To obtain the master equation for the SQUID ring dynamics, the environment is traced out to yield:
\ba
	\TD{\tilde{\rho}_{S}(t)}{t}&=&- \frac{\ui}{\hbar}\Tr_{B}([\tilde{H}_{I}(t),\tilde{\rho}_{S}(0)\otimes\DO_{B}])\\ \nonumber   &&-\frac{1}{\hbar^{2}}\int_0^t \ud s\Tr_{B}([\tilde{H}_{I}(t),[\tilde{H}_{I}(s),\tilde{\rho}_{S}		(s)\otimes \tilde{\DO}_{B}]])
\ea
For a system, linearly coupled to the environment as here, we assume a Ohmic bath with zero mean so that the first term above vanishes to give:
\bel{eq:preredfield}
	\TD{\tilde{\DO}_{S}(t)}{t}=-\frac{1}{\hbar^{2}}\int_0^t \ud s\Tr_{B}([\tilde{H}_{I}(t),[\tilde{H}_{I}(s),\tilde{\DO}_{S}(s)\otimes \tilde{\DO}_{B}]])
\ee
We note that there is increasing interest on the effect of non-linear couplings between systems (such as those with a Kerr type nonlinearity) \cite{Hu2011,Allman2010}. In such circumstances, as here, the approximations used in the standard derivation of the master equation would need to be examined in detail. The Markovian approximation further assumes that the system is only dependent on its current state and not on its state at earlier times which allows the replacement $\DO_{S}(s)\rightarrow\DO_{S}(t)$ to be applied. Substitution into~\Eq{eq:preredfield}then leads to the Redfield equation \cite{Weiss1999}:
\[
	\TD{\tilde{\DO}_{S}(t)}{t}=-\frac{1}{\hbar^{2}}\int_0^t \ud s\Tr_{B}\left\{[\tilde{H}_{I}(t),[\tilde{H}_{I}(s),\tilde{\DO}_{S}(t)\otimes \tilde{\DO}_{B}]]\right\}.
\]
The correlations with the system at different times may be made clearer by the change of variables $s= t-\tau$ where $\tau$ is interpreted as the relaxation time for the system. In the Markovian limit, memory effects must be short lived and the integrand within the dissipator decays very quickly for $\tau$ much larger than the bath correlation time. With our previous discussion of the validity of the Markovian approximation and caveats in mind, the limits of integration can therefore be extended to infinity (essentially here this requires $t \gg 1/\Omega$). This change of variable, together with interchanging the limits of integration,  gives the general form of the master equation in the interaction picture:
\beq \label{eq:7}
	\TD{\tilde{\rho}_{S}(t)}{t}=-\frac{1}{\hbar^{2}}\int_0^\infty \ud\tau\Tr_{B}\Big\{[\tilde{H}_{I}(t),[\tilde{H}_{I}(t-\tau),\tilde{\rho}_{S}(t)\otimes \tilde{\DO}_{B}]]\Big\}
\eeq
Finally, rotating these equations back into the Schr\"odinger picture yields the dynamics for the system's reduced density matrix as:
\ba \label{eq:3} 
	\TD{\DO_{S}(t)}{t}&=&-\frac{\ui}{\hbar}[\OpH_{S},\DO_{S}(t)] \\ \nonumber && -\frac{1}{\hbar^{2}}\int_0^\infty \ud\tau\Tr_{B}\Big\{\left[\OpH_{I},\left[\OpH_{I}(-\tau),\DO_{S}(t)\otimes \tilde{\DO}_{B}\right]\right]\Big\}.
 \ea
as, for linear coupling and a time-independent Hamiltonian, $\hat{\Phi}(-\tau)=e^{-{i\OpH_{S}\tau}/{\hbar}}\hat{\Phi} e^{{i\OpH_{S}\tau}/{\hbar}}$ (as $\hat{\Phi}$ commutes with $\OpH_I$ and $\OpH_B$). This  equation is of the form of a modified Liouville-von Neumann equation. The first term describes the free evolution of the system while the second term, the dissipator, represents non-unitary loss. Note that rotation to and from the interaction picture will be significantly more complex with a time-dependent external flux, or if the device dynamics includes a time-varying controller. 

Using the SQUID-environment interaction Hamiltonian above, and expanding the commutators within the integral, this can be written in the form: 

\beq \label{eq:4}\begin{split}
	\TD{\DO_{S}(t)}{t}= -\frac{\ui}{\hbar}[\OpH_{S},\DO_{S}(t)&]\\+ \frac{1}{\hbar^{2}} \int_{0}^{\infty}d\tau \Big(&\frac{i}{2}D(-\tau)[\hat{\Phi},\{\hat{\Phi}(-\tau),\DO_{S}(t)\}]\\- &\frac{1}{2}D_{1}(-\tau)[\hat{\Phi},[\hat{\Phi}(-\tau),\DO_{S}(t)]] \Big)
\end{split}\eeq
Here $\rho_{S}(t)$ describes the reduced density matrix in the external flux basis and $[\cdot]$ and $\{\cdot\}$ denote commutators and anticommtutators respectively. As the terms in the integrand of ~\Eq{eq:4} are both commutators, the cyclic property of $\Tr$ ensures that $\Tr (d\rho/dt) = 0$, thus $\Tr(\rho) = 1$ for all $t$. However Lindblad form is not assured. The functions $D$ and $D_1$ are related to the bath correlation function $B$ by \cite{Breuer2002}: 
 \beq
 D_1(-\tau) + \ui D(-\tau) = 2 \EX{BB(-\tau)}_B
 \eeq
where the expectation value with respect to the bath is given by $\EX{BB(-\tau)}_B = \Tr_B\{BB(-\tau)\DO_B\}$. In this case, the coupling constants, $\kappa_n$, in~\Eq{eq:1} are determined by a quasi-continuous spectral density $J(\omega)$, which describes the absorption and emission of energy arising from the coupling to the environment. The dissipation and noise kernels can be written in terms of the spectral density as \cite{Breuer2002}: 
\beq \begin{split} \label{eq:23}
	D(-\tau)&= 2\hbar \int_{0}^{\infty}d\omega J(\omega) \sin{( \omega \tau)} \\ D_{1}(-\tau) &=  2\hbar \int_{0}^{\infty}d\omega J(\omega)\coth{\left(\frac{\hbar\omega}{2k_{B}T}\right)} \cos{( \omega \tau)}
\end{split} \eeq
Whilst the first expression is easy to evalutate for an Ohmic bath, the second requires the separation into slowly and rapidly oscillating terms, as indicated in [\onlinecite{Gao1997}], which enables us to write $D_1(-\tau)$ as approximately:
 \beq
 D_1(-\tau) = \frac{\omega_0}{2} \coth{\left(\frac{\hbar\omega_0}{2k_{B}T}\right)}\int^\infty_0  d\omega\frac{J(\omega)}{\omega} \cos{( \omega \tau)}
 \eeq
For an Ohmic bath with a Lorentz-Drude cutoff function, with cutoff frequency $\Omega$, the spectral density is given by:
\beq 
J(\omega)= \frac{2C\gamma}{\pi}\omega \frac{\Omega^{2}}{\Omega^{2}+\omega^{2}}
\eeq
where  $\omega$ is a bath frequency and $\gamma$ represents the damping rate of the system.

In this case, the dissipation \cite{Breuer2002} and noise \cite{Gao1997} kernels, $D(-\tau)$ and $D_{1}(-\tau)$, may be written, respectively, as:
\[
\begin{split}
	D(-\tau)&= 2C\gamma\hbar\Omega^{2}e^{-\Omega |\tau|} \sgn{\tau}
 \\ D_{1}(-\tau) &=  C\hbar\gamma\Omega\omega_{0} \coth{ \left(\frac{\hbar\omega_{0}}{4k_{B}T} \right)}e^{-\Omega|\tau|}
\end{split}
\]
in the mid-low temperature regime \cite{Breuer2002, Gao1997}, for system thermal energy $k_{B}T$. In the limit temperature $T \rightarrow 0$ the noise kernel reduces further to
\beq 
\begin{split} \label{eq:5}
	D_{1}(-\tau) &=  C\hbar\gamma\Omega\omega_{0} e^{-\Omega|\tau|}
\end{split}
\eeq
The approximation used in Eq. (16) has an easier justification at higher temperatures. At low temperatures it would be more accurate to swap the order of the time integral in Eq. (13) and the frequency integral in Eq. (15), as is done for the special case of the Quantum Brownian Motion\cite{Gardiner1991,Breuer2002, Schloss2007,Caldeira1983,Chou2008, Diosi1993, Halliwell1996, Hu1992, Unruh1989, Massignan2015, Fleming2011, Mounier1998}. Details of this will be presented in a future work.

\section{Integrating the Master Equation \label{IME}}

An issue which arises with QBM is a logarithmic cut-off divergence (leading to a $\log(\Omega)$ dependence in the diffusion terms) in the exact solution of the master equation, thus making the large $\Omega$ limit difficult. Most approximations stop at first order in $\omega_0/\Omega$, before the log-term enters, and this rather begs the question of how accurate this is and consequently we seek here both first and second order solutions. To derive a useful master equation it is necessary to evaluate, or at least approximate, the dissipator integral in  (\ref{eq:4}). A common means of approximating  the relaxation-time dependent flux term $\hat{\Phi}(-\tau)$ is through a power series expansion in $\tau$, such that:
\bel{eq:series}
\hat{\Phi}(-\tau)= \sum_{n} A_{n}[\hat{\Phi}] \tau^{n} 
\eeq
where the functional $A_{n}[\hat{\Phi}]$ is found by equating powers of $\tau$ from the Baker-Campbell-Hausdorff expansion of $\hat{\Phi}(-\tau)=e^{-{i\OpH_{S}\tau}/{\hbar}}\hat{\Phi} e^{{i\OpH_{S}\tau}/{\hbar}}$ i.e.:
\beq \begin{split} \label{eq:BCH} 
\hat{\Phi}(-\tau) = \hat{\Phi} &+ \tau \left[-\frac{i\OpH_{S}}{\hbar}, \hat{\Phi}\right]+ \frac{\tau^2}{2!}\left[-\frac{i\OpH_{S}}{\hbar},\left[-\frac{i\OpH_{S}}{\hbar},\hat{\Phi}\right]\right]
\\ &+ \cdots+ \frac{\tau^n}{n!}\left[-\frac{i\OpH_{S}}{\hbar},...,\left[-\frac{i\OpH_{S}}{\hbar},\hat{\Phi}\right]\right]
\end{split}\eeq
For the simpler case of a quantum Brownian particle in a harmonic oscillator potential, each of the $A_{n}[\hat{\Phi}]$ is proportional to either the position or momentum operator, with pre-factors which add to give trigonometric terms \cite{Massignan2015}. Unfortunately the same cannot be said for the SQUID. Due to the nonlinear nature of the Josephson junction term in the Hamiltonian, the series grows in complexity as the order is increased. For this reason it is not possible to evaluate $\hat{\Phi}(-\tau)$ analytically and it is necessary to truncate the series in ~\Eq{eq:series}. Analysis of this series shows it to be convergent and a more detailed study will follow in later work. Including more terms in the series though should lead to increasingly accurate master equations and here we explore the impact of truncating to first and second order.  Note that if the system Hamiltonian were to be time dependent (possess a time dependent external flux $\Phi_{x}(t)$), the series would grow significantly in complexity and this method may not be applicable.

Substituting~\Eq{eq:series} into the expressions for the dissipator of~\Eq{eq:4} yields the non-Lindblad master equation:
\begin{widetext}
\beq  
	\TD{\DO_{S}(t)}{t} 
	= -\frac{\ui}{\hbar}[\OpH_{S},\DO_{S}(t)]
	+
	\frac{\ui C\gamma\Omega}{\hbar}
	\COM{\Phi}{\ACOM{\sum_{n}\frac{n!}{\Omega^{n}}A_{n}[\hat{\Phi}]}{\DO_S(t)}}
	-\frac{C\hbar\gamma\omega_{0}}{2\hbar} 
	\COM{\Phi}{\COM{\sum_{n} \frac{n!}{\Omega^{n}} A_{n}[\hat{\Phi}]}{\DO_S(t)}}  \label{eq:meq_sums}
	\eeq
\end{widetext}
 where the identities for the dissipation and noise terms:
\begin{widetext}
\beq \label{eq:8} \begin{split}
\frac{i}{2\hbar^{2}}\int_{0}^{\infty} \ud\tau D(-\tau) \hat{\Phi}(-\tau) &= \sum_{n}\frac{iC\gamma\Omega}{\hbar}\frac{n!}{\Omega^{n}}A_{n}[\hat{\Phi}]
\\
-\frac{1}{2\hbar^{2}}\int_{0}^{\infty} \ud \tau D_{1}(-\tau)\hat{\Phi}(-\tau) &=
 -\frac{C\hbar\gamma\omega_{0}}{2\hbar} 
 \sum_{n} \frac{n!}{\Omega^{n}}A_{n}[\hat{\Phi}]
\end{split}\eeq
\end{widetext}
have been used alongside the identity $\Omega^{n+1} \int_{0}^{\infty} \ud\tau \tau^{n} e^{-\Omega \tau} = n!$. 

\section{First Order Master Equation}
If the series of~\Eq{eq:series} is truncated to first order in $\tau$ then the summations in~\Eq{eq:meq_sums} can be simplified accordingly:
\beq \label{eq:firstordersum}
\sum_{n} \frac{n!}{\Omega^{n}}A_{n} \approx A_{0} + \frac{1}{\Omega} A_{1} 
= \hat{\Phi}  -\frac{\hat{Q}}{\Omega C}
\eeq
so that~\Eq{eq:meq_sums} yields the first order master equation:
\beq \begin{split} \label{eq:meqFirstOrderOne}
\TD{\DO_S}{t}= &-\frac{\ui}{\hbar}[\OpH_{S},\DO_S] +
\overbrace{\frac{\ui C\gamma\Omega}{\hbar}[\hat{\Phi}^{2},\DO_S]}^{\text{renormalises $L$}}
-\overbrace{\frac{\ui \gamma}{\hbar}[\hat{\Phi},\{\hat{Q},\DO_S\}]}^{\text{{dissipation term}}}
\\& -
\frac{C\omega_{0} \gamma}{2\hbar}\bigg(\underbrace{[\hat{\Phi},[\hat{\Phi},\DO_S]]}_\text{noise term} - \underbrace{\frac{1}{\Omega C} [\hat{\Phi}, [\hat{Q},\DO_S]]}_{\text{
first order cutoff
}}\bigg)
\end{split} \eeq
It is worth remarking, at this stage, that additional capacitive coupling in the interaction Hamiltonian (Eq. 2) will lead to a much more complicated expression than ~\Eq{eq:meqFirstOrderOne} due the the presence of a commutation relation between the charge operator and the Josephson coupling energy and would inevitably lead to a non-linear dependence on external flux even in a first order master equation. We believe this would produce noticeable differences in theory which could be observed experimentally even for modest couplings (again a more detailed study will be the subject of future work). The final term in eqn (25) vanishes in the limit of high cutoff frequency. This limit is often assumed in quantum optics and the term neglected but, as indicated above, is not be applicable to condensed matter systems and we retain it for this reason, and because it is also a necessary ingredient for turning ~\Eq{eq:meqFirstOrderOne} into Lindblad form.

The second term in \Eq{eq:meqFirstOrderOne} is simply a renormalisation of the potential, or more specifically a shift in the SQUID inductance \cite{Prance1991,Diggins1994, Spiller1991} by a factor of $\lambda = \frac{2\Omega\gamma}{\omega^{2}_{0}(1+ 2\Omega\gamma /\omega^{2}_{0})}$ and can therefore be absorbed into the free evolution part of the equation to give:
\beq \begin{split} \label{eq:13}
\TD{\DO}{t}= &-\frac{\ui}{\hbar}[H_{S_1},\DO] -\frac{\ui\gamma}{\hbar}[\hat{\Phi},\{\hat{Q},\DO\}] \\& -\frac{C\omega_{0} \gamma}{2\hbar}\bigg([\hat{\Phi},[\hat{\Phi},\DO]] - \frac{1}{\Omega C} [\hat{\Phi}, [\hat{Q},\DO]]\bigg)
\end{split} \eeq 
where $\OpH_{S_1}$ is of exactly the same form as $\OpH_{S}$ as in~\Eq{eq:HamFluxBasis} but uses the bare inductance of the SQUID ring, since $L_0 = L/(1-\lambda)$, instead of $L$. \Eq{eq:13} is a Caldeira-Leggett equation \cite{Caldeira1983}, rather than in the Lindblad form  of~\Eq{eq:meqLindblad}, and thus does not ensure all solutions will be physically sensible\cite{Munro1996} (i.e. a density operator that is postive). The simplest way to address this issue is to transform~\Eq{eq:13} into Lindblad form, as for QBM \cite{Breuer2002, Gao1997, Wiseman1998}. This is achieved through the addition of a term proportional to $[\OpQ,[\OpQ, \DO]]$ . The physical significance of this addition becomes clear when considering the same system capacitively (rather than inductively) coupled to the bath, when such a term arises naturally. One can then think of this addition as the inclusion of a capacitive element in the interaction, an effect that will be presented in future work. It should be noted that unlike the case of QBM at high temperatures, the additional term is not necessarily small. Nevertheless, proceeding this way leads to 
\beq \begin{split}\label{eq:meqLindbladFirst}
\TD{\DO}{t} &= -\frac{\ui}{\hbar}[\OpH,\DO] + \frac{1}{2}\left([\OpL, \DO \OpL^{\dagger}] + [\OpL\DO, \OpL^{\dagger}]\right)
\\
\OpH &=\OpH_{S_1}+\frac{\hbar \gamma}{2}\left(\OpX \OpP + \OpP \OpX  \right)
\\
\OpL &=\gamma^\Half\left[\OpX + \left(\ui - \frac{\xi}{2}\right) \OpP \right]
\end{split}
\eeq
where we have introduced the dimensionless quantities $\OpX=\sqrt{\frac{C\omega_0 }{\hbar}}\Flux$, $\OpP=\sqrt{\frac{1}{C \hbar \omega_0}}  \Charge$ and $\xi=\omega_0/\Omega$. 
There are a number of observations to be made here in relation to the introduction of the $[\OpQ,[\OpQ, \DO]]$ into \Eq{eq:13}. First \Eq{eq:meqLindbladFirst} recovers, in the limit $\xi\rightarrow 0$, a more familiar Lindblad proportional to the annihilation operator. What the derivation here demonstrates is that in assuming $\OpL =\sqrt{2\gamma} \Opa$, for some $\gamma$, a significant adjustment to the master equation is being made. Second, it is clear that, within the Hamiltonian $\OpH$, there exists a squeezing term, which cannot be included in the Lindblad terms, but which may instead be included in the system Hamiltonian. This arises as a corollary of applying the Lindblad process and its inclusion is very often neglected in the literature. However, it is a necessary part of the system evolution which provides a physical frequency shift, and is essential in recovering the quantum to classical transition~\cite{Everitt2009a, Everitt2009, Montina2008,Kapulkin2008,Li2011}. 

This is evident from the harmonic oscillator component of the SQUID ring Hamiltonian, $\OpH  = \OpH_{S_1}+(\hbar \gamma)/2\left(\OpX \OpP + \OpP \OpX  \right)= \hbar\omega(\hat{a}^{\dagger}\hat{a}+1/2) + (\hbar\gamma i)/2\left(\hat{a}^{\dagger 2}-\hat{a}^{2}\right)$; the significance of the second (the squeezing) term appears when considering the correspondence limit. If the quantity $\Tr\left(\frac{d}{dt}(\rho\hat{a})\right)$ is found from ~\Eq{eq:meqLindblad}, without the squeezing term, one obtains an expression for the expectation value of the evolution of the position operator:
\beq \label{eq:7}
\EX{\hat{x}(t)} = \left(\EX{\hat{x}_{+}}e^{i\omega t} + \EX{\hat{x}_{-}}e^{-i\omega t}\right)e^{-\gamma t}
\eeq
which describes a system oscillating at a frequency $\omega$ and decaying at a rate $e^{-\gamma t}$.

Although there appears to be agreement with decay rate in classical models for the damped harmonic oscillator, frequency shifts are not accounted for; and this violates the correspondence principle. This result suggests two things: Lindblad operators describe dissipation only, and the frequency shift is described by the additional Hamiltonian term. The impact of the squeezing term can be seen by performing a Bogoliubov transform \cite{Bogoliubov1947} so that the Hamiltonian may be written in terms of a new set of raising and lowering operators, $\hat{b}^{\dagger}$ and $\hat{b}$
\[\begin{split}
\hat{b} &= u \hat{a} + v\hat{a}^{\dagger} \\ \hat{b}^{\dagger}&= u^{*}\hat{a}^{\dagger} + v^{*}\hat{a}
\end{split}\]
that reproduce $\hat{a}^{\dagger}$ and $\hat{a}$ in the limit where $\gamma \rightarrow 0 $. Satisfying the requirement that $|u|^{2}+|v|^{2} =1$ through the assumption that the constants $u=\sec{\theta}$ and $v=i\tan{\theta}$, allows the Hamiltonian to be rewritten as
\[
H'  = \hbar\tilde{\omega}\hat{b}^{\dagger}\hat{b} = \hbar\omega\sqrt{1-\frac{\gamma^{2}}{\omega^{2}}}\hat{b}^{\dagger}\hat{b}
\]
It is clear to see that this term is responsible for the frequency shift of the dissipating system.

\begin{figure}[!t]
\includegraphics[width=\linewidth]{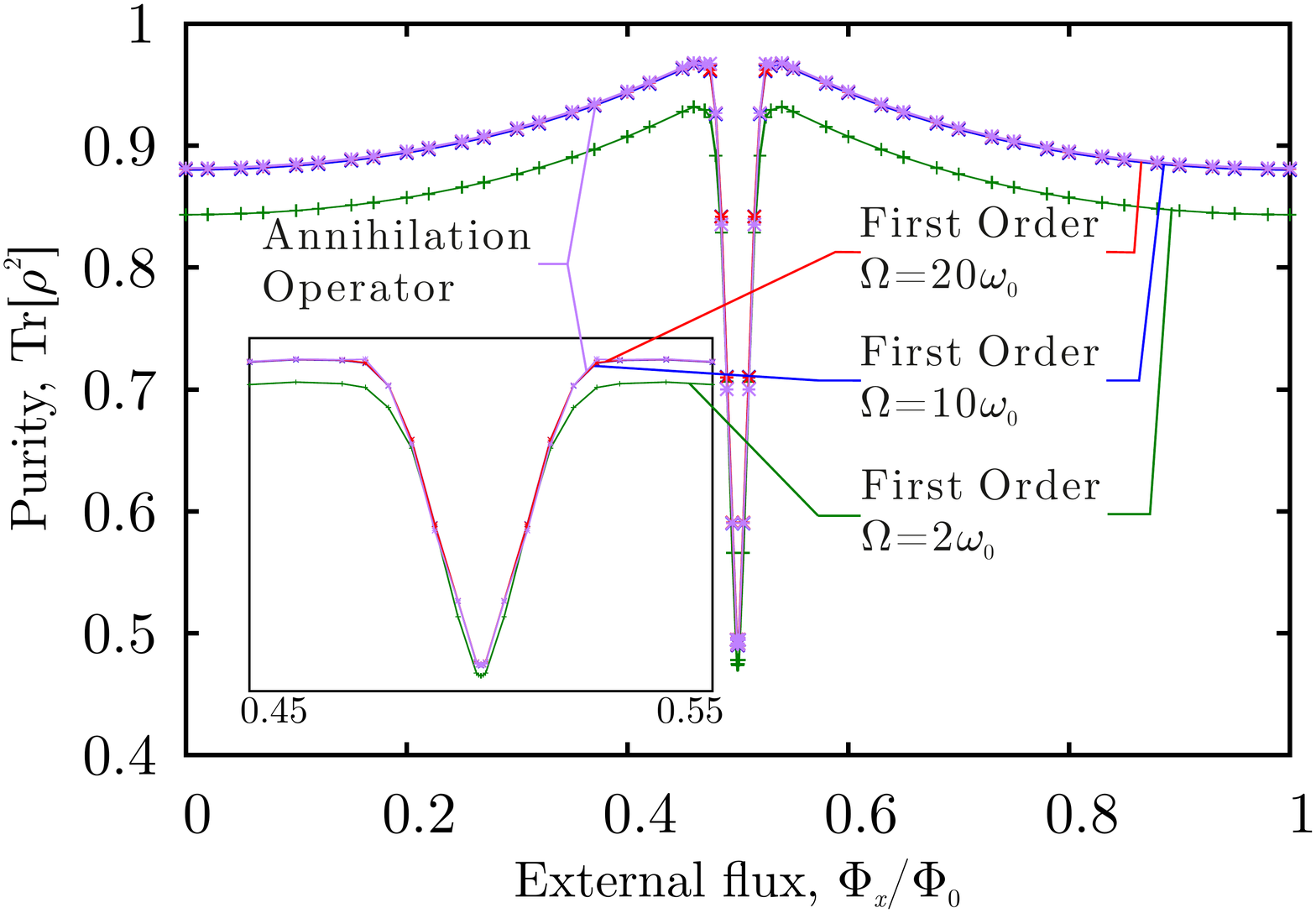}
\caption{
(colour online) To quantify the importance of cutoff frequency, $\Omega$, in the first order master equation~\Eq{eq:meqLindbladFirst}, we show $\Trace{\DO^2}$ as a function of external flux $\Flux_x$ for the steady state solution to the master equation for $\xi=\omega_0/\Omega$ equal to $0\ (\Omega=\infty)$, $0.05\ (\Omega=20\omega_0)$, $0.1\ (\Omega=10\omega_0)$ and $0.5\ (\Omega=2\omega_0)$. We see that $\Omega=\infty$ and $\Omega=20\omega_0$ are indestinguishable whilst near the dip at $\Phi_x=0.5\Phi_0$ there are small differences at $\Omega=10\omega_0$. 
While the functional form is similar the effect of cut-off frequency is significant for $\Omega=2\omega_0$.
Note, circuit parameters are $C=5\times10^{-15}$F, $L=3\times 10^{-10}$H and $I_c\approx 3\mu\mathrm{A}$.
The sharp dip at $\Phi_x = \Phi_0/2$ is due to the fact that the SQUID's potential becomes a double well and the ground energy eignstate is a Schr\"odinger cat (i.e. a macroscopically distinct superposition of states). Decoherence of this state produces a statistical mixture of states equally localised in each well - as there is a 50\% chance of being in either well $\Trace{\DO^2}=0.5$ at $\Phi_x = \Phi_0/2$. 
As we move away from this bias point the ground state rapidly loses its Schr\"odinger cat structure and so decoherence is less significant at these values. The width of this dip is related to the the barrier hight and can be changed by altering circuit parameters.\label{fig:firstorder} 
}
\end{figure}
The Lindblad in ~\Eq{eq:meqLindbladFirst} is a function of cut-off frequency as $\xi=\omega_0/\Omega$, we now establish how significant this is when compared with simply assuming a Lindblad term proportional to the annihilation operator. 
There are many ways that we can quantify the effect of changing cut-off frequency, but as our focus in this work is on estimating the effects of environmental decoherence we choose to compare the purity, $\Trace{\rho^2}$, of the steady state solution to ~\Eq{eq:meqLindbladFirst} as a function of external flux and cut-off frequency. This is shown in ~\Fig{fig:firstorder}. 
We first note that in the limit $\Omega \rightarrow \infty$ we have $\xi\rightarrow 0$ and the Lindblad reduces to the annihilation operator times $\sqrt{2\gamma}$ and the standard form of the master equation that has been applied to SQUIDs in previous work \cite{Everitt2014,Viola1997,Nielsen2000,Dajka2009,Nakano2004}. 

In this work, we have chosen reasonable SQUID parameters values of $C=5\times10^{-15}$F and $L=3\times 10^{-10}$H are used in all caluculations together with a Josephson coupling energy \cite{Everitt2001a} of $\hbar\nu =I_c \Phi_0/2\pi =9.99\times10^{-22}$J, where $\Phi_{0} = {h}/{2e}$ is the flux quantum and $I_c$ is the critical current of the weak link (here $I_c\approx 3\mu\mathrm{A}$). The external environment is defined by the parameters $\gamma$, $\Omega$, $\hbar \nu$, and $\Phi_{x}$ where the damping rate $\gamma$ determines the rate of loss in the system. Treating the environment as a cavity of harmonic oscillator modes, this loss is directly proportional to the cavity quality factor $Q_{c} = 2\pi \omega_{c}/\gamma$ for cavity frequency $\omega_{c}$. This quality factor can range from $Q_{c} \sim10^{2}$ to $Q_{c} \sim10^{6}$ or higher \cite{Liu2005, Whittaker2014}. The cutoff frequency, $\Omega$, defines the peak frequency of the bath's spectral density which has a similar form to the impedance in Josephson circuits \cite{Makhlin2004, Shnirman2002}. 
The results shown in~\Fig{fig:firstorder} might lead us to conclude that for a cut off frequency of $\Omega=10\omega_0\ (\xi = 0.1)$ and higher (lower) that the usual choice of a Lindblad proportional to the annihilation operator is a good one. In the next section we show that this conclusion is incorrect.

\section{Second Order Approximation}
Although for systems of this type it is often assumed to be adequate, truncation at first order of series (\Eq{eq:firstordersum}) may not always suffice and higher order terms in $\tau$ (or equivalently $\omega_0/\Omega$) may be important; consideration of a second order expression will help to justify that. It is also important to explore the impact of higher order terms as higher order models may differ quantitatively, if not qualitatively, to the first order model. Expanding \Eq{eq:firstordersum} to second order in $\tau$ we obtain:
\beq \label{eq:15} 
\sum_{n} \frac{n!}{\Omega^{n}}A_{n} \approx  \Flux  -\frac{\Charge}{\Omega C} - \frac{\omega^{2}}{\Omega^{2}}\left(\Flux + \frac{2\pi \hbar \nu L}{\Phi_{0}}\sin{\left(\frac{2\pi}{\Phi_{0}}(\Flux+\Phi_{x})\right)}\right)
\eeq%
where the external flux dependence, originating from the non-linear SQUID potential, can be seen to enter the dissipator for the first time. Substituting (\ref{eq:15}) into (\ref{eq:8}) then allows (\ref{eq:4}) to be rewritten as:
\begin{widetext}
\beq \begin{split} \label{eq:meqSecondPreLindblad}
	\TD{\DO_{S}}{t}&= -\frac{\ui}{\hbar}[\OpH_{S},\DO_{S}(t)] 
	+\frac{\ui \gamma \Omega C  }{\hbar} 
	\Bigg(
	\overbrace{\left(1-\frac{\omega_0^{2}}{\Omega^{2}}\right)[\Flux^2,\DO_{S}(t)]]}^{\text{renormalises $L$}}
	-
\overbrace{
	\frac{1}{\Omega C}[\Flux,\{\Charge,\DO_{S}(t)\}]
	}^{\text{1\St order dissipation}}
	-
	\overbrace{
\frac{2\pi \hbar\nu L}{\Phi_{0}}\frac{\omega_0^{2}}{\Omega^{2}}\left[\Flux,\left\{\sin{\left(\frac{2\pi}{\Phi_{0}}(\Flux+\Phi_{x})\right)},\DO_{S}(t)\right\}\right]}
^{\text{2\Nd order dissipation}}
	\Bigg)
	\\ &
	-\frac{\gamma \omega_0 C  }{2\hbar}
	\Bigg(
	\underbrace{\left(1-\frac{\omega_0^{2}}{\Omega^{2}}\right)[\Flux,[\Flux,\DO_{S}(t)]]}_\text{1\St and 2\Nd order noise}
	 -
	 \underbrace{\frac{1}{\Omega C}[\Flux,[\Charge,\DO_{S}(t)]]}
	 _\text{1\St order in cutoff}
	-
	\underbrace{
	\frac{2\pi \hbar\nu L}{\Phi_{0}}\frac{\omega^{2}_0}{\Omega^{2}}\left[\hat{\Phi},\left[\sin{\left(\frac{2\pi}{\Phi_{0}}(\hat{\Phi}+\Phi_{x})\right)},\DO_{S}(t)\right]\right]}
	_\text{2\Nd order cutoff}
	\Bigg)
\end{split}\eeq
\end{widetext}
where once again $\OpH_{S}$ consists of the true inductance of the SQUID ring after second order renormalisation  is accounted for, i.e. $\lambda = \left(2\gamma\Omega\left(1-\frac{\omega_{0}^{2}}{\Omega^{2}}\right)\right)/\omega_{0}^{2}\left(1+\frac{2\gamma \Omega}{\omega_{0}^{2}}\left(1-\frac{\omega^{2}_{0}}{\Omega^{2}}\right)\right)$. 

\begin{figure}[!t]
\includegraphics[width=\linewidth]{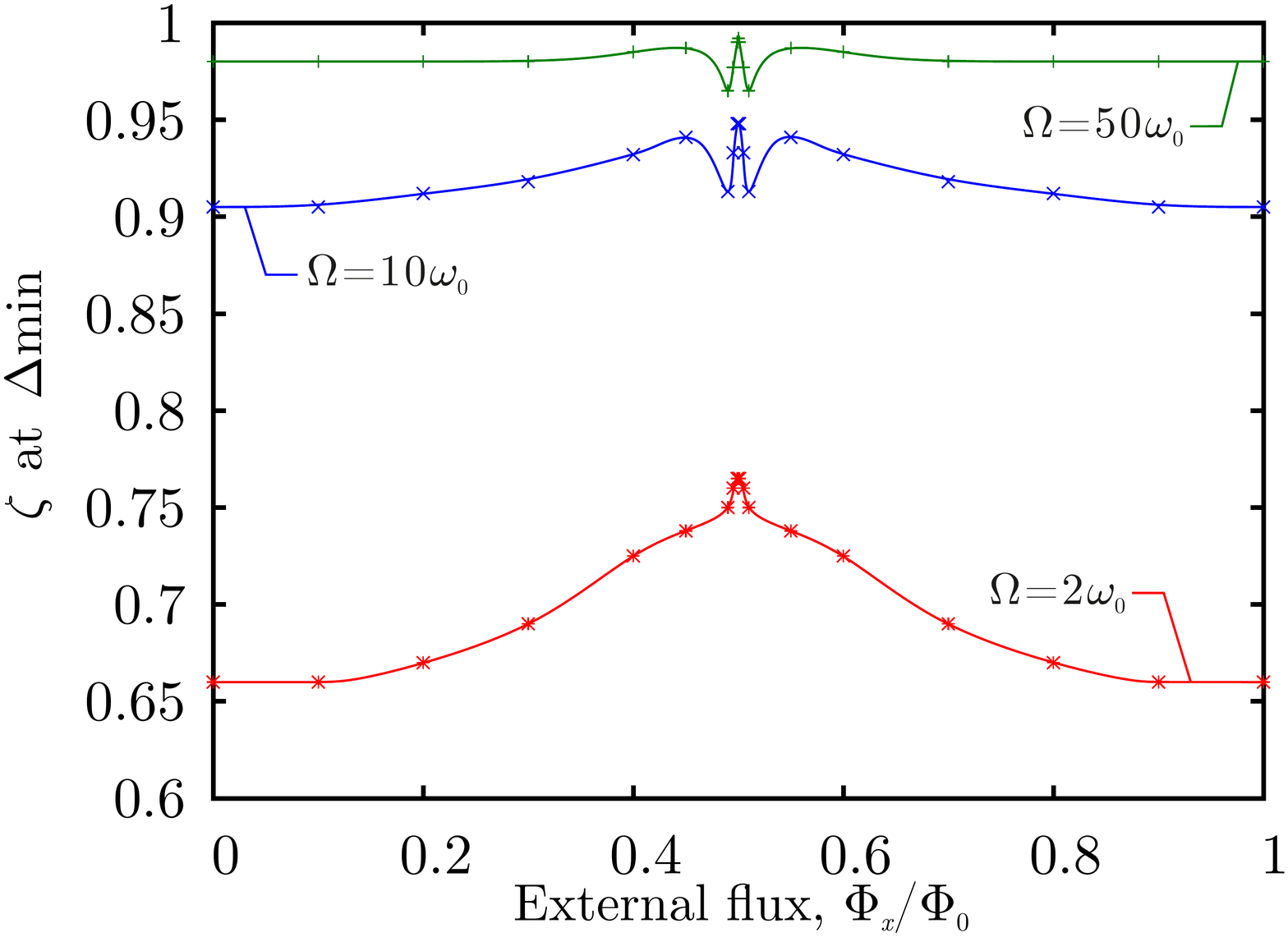}
\caption{
(colour online) A plot of the second order weighting parameter $\zeta$ that minimises the difference between first and second order master equations as a function of external flux. $\Delta_\mathrm{min}$ is defined to be the minimal difference in steady state purity between first and second order models, for system parameters $\Omega, \Phi_{x}$. We see that the $\zeta$ that is minimally invasive is a non-linear function of external flux. For high cut off frequency this is approximated by $\zeta = 1 - \omega_0/\Omega$ (where we note that this approximation is less good around $\Phi_{x}=\Phi_{0}/2$). \label{fig:zeta} 
}
\end{figure}

\Eq{eq:meqSecondPreLindblad} is once again not of Lindblad form, and suffers from the associated problems. However it may be made so by following the same process as in the first order case, and it is of interest to observe the form that the Lindblad operators now take. Two Lindblads $\OpL_{1} = \alpha_{1} \hat{\Phi} + \epsilon_{1} \hat{Q}$ and $\OpL_{2} = \alpha_{2} \hat{\Phi} + \epsilon_{2} \sin{\left(\frac{2\pi}{\Phi_{0}}(\hat{\Phi}+\Phi_{x})\right)}$  are needed; the first is an annihilator while the second represents a correction to the environmental interactions and is a function of the external flux control parameter $\Phi_x$.  

There is some flexibility to the manner in which the fifth term in ~\Eq{eq:meqSecondPreLindblad} may be split between the two Lindblads, $\OpL_1$ and $\OpL_2$. The weighting of this split, with respect to first and second order contributions, is characterised in this work by the weighting parameter $\zeta$ and is allocated in such a way that $-(1-\zeta)\frac{\gamma \omega_0 C  }{2\hbar}\left(1-\frac{\omega_0^{2}}{\Omega^{2}}\right)[\Flux,[\Flux,\DO_{S}(t)]]$ contributes to $\OpL_1$ and $-\zeta\frac{\gamma \omega_0 C  }{2\hbar}\left(1-\frac{\omega_0^{2}}{\Omega^{2}}\right)[\Flux,[\Flux,\DO_{S}(t)]]$ contributes to $\OpL_2$. Usually a `minimally invasive' approach is taken to ensure that first order terms remain dominant and the extra term needed for $\OpL_2$ is as small as possible. In~\Fig{fig:zeta} we show the value of $\zeta$ which finds the minimum difference
$
\Delta_\mathrm{min}
$
in steady state purity between the first and second order master equations. For most systems this would be expected to be constant value but for the SQUID ring it is non-linearly dependent on external flux.  This is not as surprising as it might first seem as SQUID rings are known to effect externally coupled oscillators (tank-circuits) in a non-linear way and the environment is considered as an infinite bath of such oscillators. As a result we expect that the modelling process should also yield results that are also non-linearly dependent on the external flux. 

It is therefore the case that the Lindblad form of the master equation expressed to second order should contain a correction that is dependent on external flux and cutoff frequency -- $\zeta(\Omega,\Phi_x)$. These and some other subtleties will be explored in a followup work. We see that for high cut-off frequency that choosing $\zeta = 1 - \omega_0/\Omega = 1-\xi$ is a good approximation to a minimally invasive master equation (especially away from $\Phi_x=0.5$). With this choice, the Lindblad operator $\OpL_1$ again approaches the annihilation operator in the high cut off limit, where $\xi \rightarrow 0$. In the remainder of this work we will thefore make the approximation that $\zeta = 1 - \omega_0/\Omega$.
Within this model, frequency shifts are still accounted for, as they are enclosed within the third term in \Eq{eq:meqSecondPreLindblad}. The second order equation also possesses a second frequency shift. It must be expected that higher order approximations in $\omega_0/ \Omega$ will introduce additional Lindblad operators and additional frequency renormalisation, this again will be investigated in future work.

If the additional terms, required to bring the equation into Lindblad form are included in~\Eq{eq:meqSecondPreLindblad}, one obtains:
\begin{widetext}
\beq \begin{split}\label{eq:meqLindbladSecond}
\TD{\DO}{t} &= -\frac{\ui}{\hbar}[\OpH,\DO] + \frac{1}{2}\sum_j\left([\OpL_j, \DO \OpL_j^{\dagger}] + [\OpL_j\DO, \OpL_j^{\dagger}]\right)
\\
\OpH &=\OpH_{S_2}
+\frac{\hbar \gamma}{2}\left(\OpX \OpP + \OpP \OpX  \right)
+\sqrt{\beta \xi \frac{\nu}{\Omega}} \OpX \sin{\left(\sqrt{\beta \omega_0\over \nu}\OpX + 2\pi \frac{\Phi_x}{\Phi_0}\right)}
\\
\OpL_1 &=\gamma^\Half\left[\sqrt{\left(1-\xi\right)\left(1-\xi^2\right)}\OpX + \left(\ui - \frac{\xi}{2}\right)\sqrt{\frac{1}{\left(1-\xi\right)\left(1-\xi^2\right)}}  \OpP\right]
\\
\OpL_2 &=
\gamma^\Half 
\Bigg[\sqrt{\xi\left(1-\xi^2 \right)}\OpX + 
\sqrt{\frac{\xi}{\left(1-\xi^2 \right)}}\left(\ui - \frac{\xi}{2}\right) \sqrt{\beta\frac{\nu}{\omega_0}}\sin{\left(\sqrt{\beta \omega_0\over \nu}\OpX + 2\pi \frac{\Phi_x}{\Phi_0}\right)}\Bigg]
\end{split}
\eeq
\end{widetext}
where here we introduced the parameter $\beta=2\pi L I_c/\Phi_0$, related to the critical current $I_c=2\pi \hbar \nu /\Phi_0$, which is frequently used in semi-classical analysis to separate hysteretic ($\beta >1$) from non-hysteretic behaviour ($\beta \le 1$).

\begin{figure}[!tb]
\includegraphics[width=\linewidth]{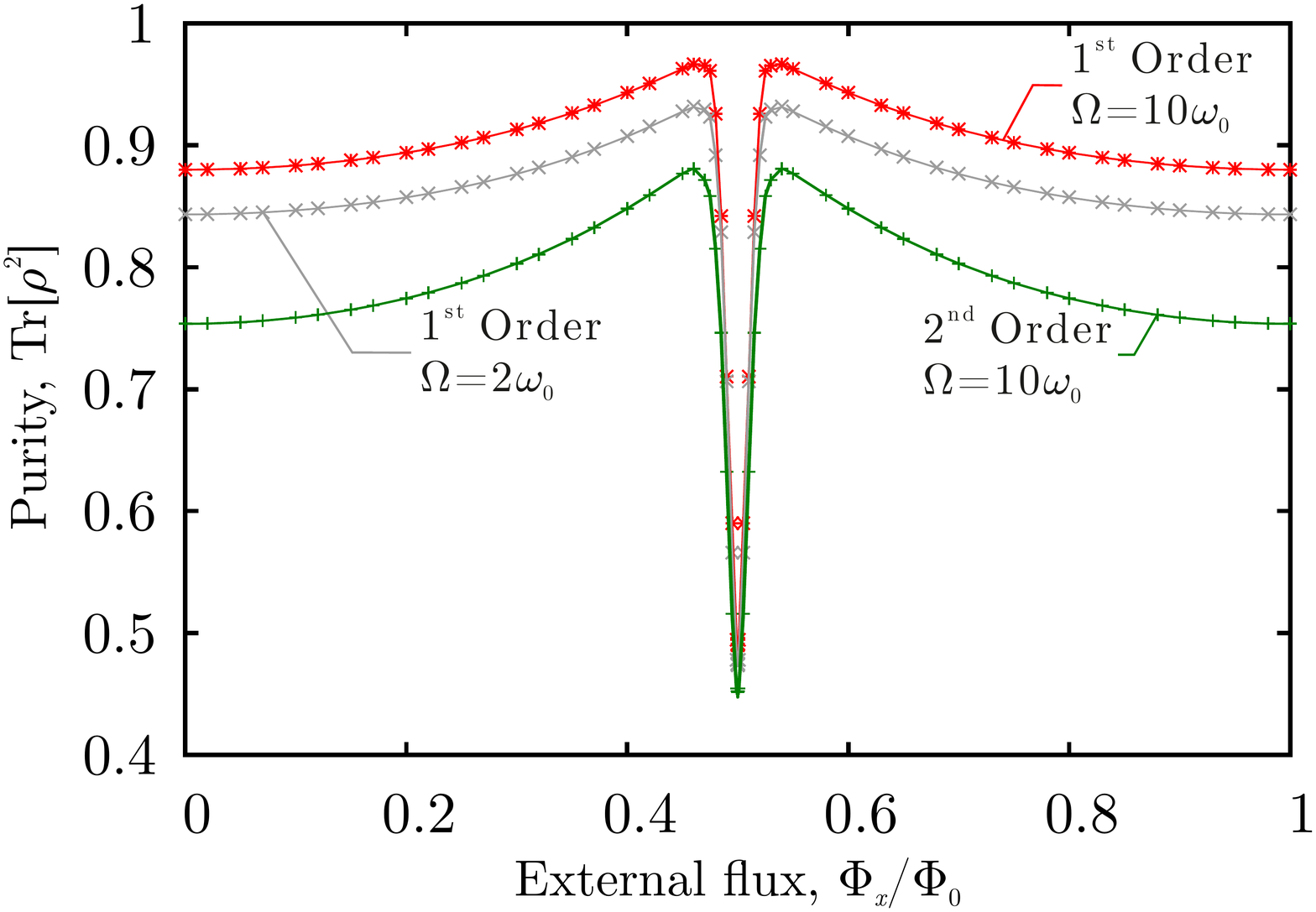}
\caption{
(colour online) The purity $\Tr\{\DO^{2}(t)\}$ of the steady state solutions of the first order, \Eq{eq:meqLindbladFirst}, and second order~\Eq{eq:meqLindbladSecond}, Lindblad master equations. 
In this figure we see evidence that the order of truncation has a bigger effect on the steady state purity than one might expect when compared to that of decreasing cut-off frequency.  \label{fig:secondorder} 
}
\end{figure}

In ~\Fig{fig:secondorder} we compare the purity $\Tr\{\DO^{2}(t)\}$ of the steady state solutions of the first order, \Eq{eq:meqLindbladFirst}, and second order~\Eq{eq:meqLindbladSecond}, Linblad master equations for a cut-off frequency of $\Omega=10\omega_0$. We have also included for comparison  the first order master equation steady state purity for $\Omega=2\omega_0$. In~\Fig{fig:firstorder},  for a cut-off frequency of $\Omega=10\omega_0$, we concluded that there was little difference between the steady state solution to the first order corrected master equation and one that just assumed an annihilation operator as a Lindblad. In~\Fig{fig:secondorder}, for the same value of cut-off frequency, we observe that the steady state purity is much lower and changes slightly in functional form in the second order model. 
This indicates that neither the annihilation operator nor first order Lindblads are sufficient to quantitatively model the effects of decoherence on the SQUID ring.

\begin{figure}[!t]
\includegraphics[width=\linewidth]{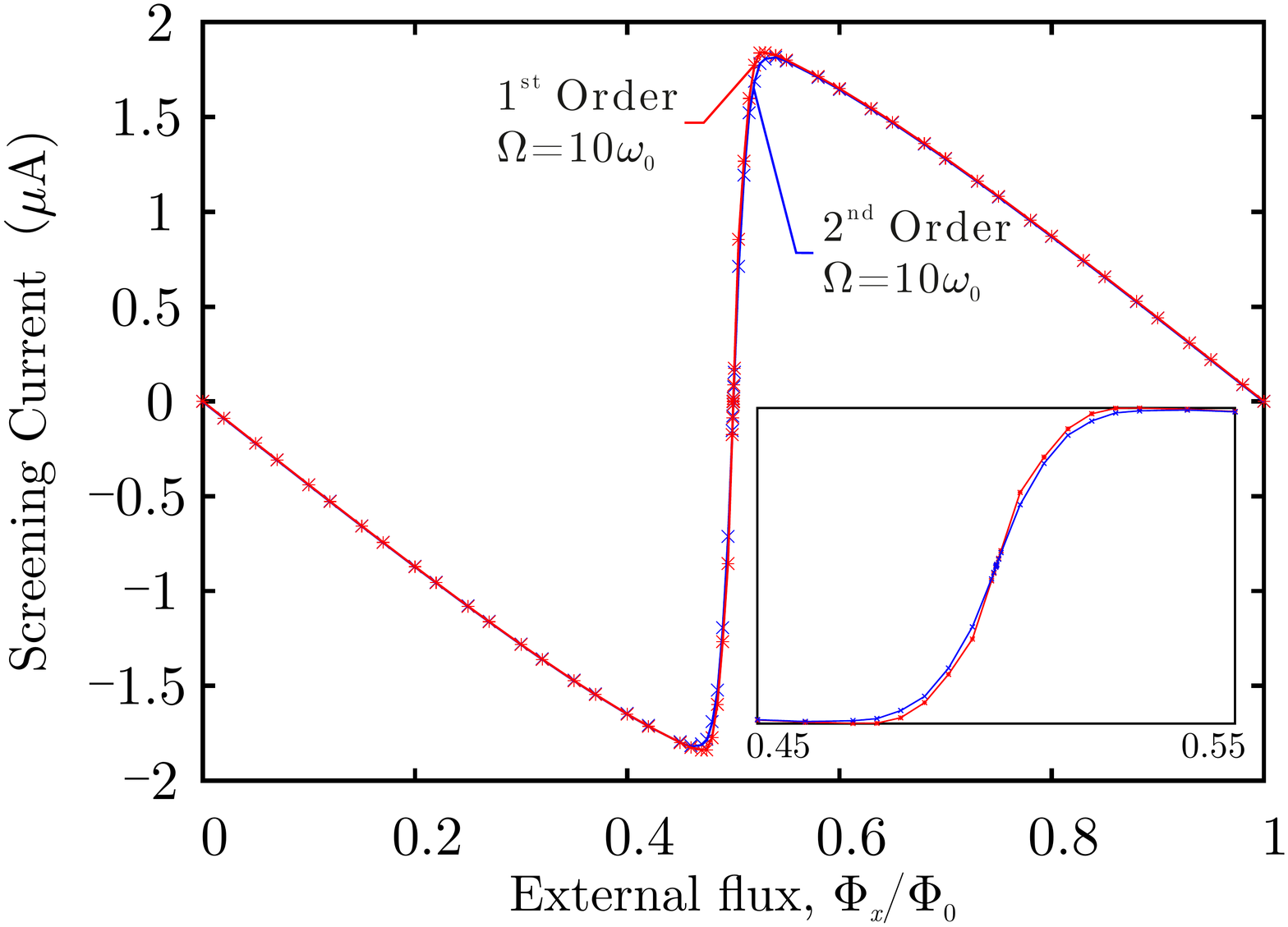}
\caption{
(colour online) A plot of the expectation value of screening current, $\EX{\Flux/ L}$ as a function of external flux for first order (red) and second order (blue) models at a bath cutoff frequency of $\Omega = 10\omega_0$. Despite the two models differing quite largely in terms of steady state purity, the expectation values of observables remain very similar. \label{fig:screeningcurrent}
}
\end{figure}

The difference between first and second order models is less obvious when considering the expectation value of observables, such as screening current, as shown in \Fig{fig:screeningcurrent}. This suggests that device characterisation based solely on simple expectation values of observables such a flux may not be sufficient and a more rigorous analysis of decoherence times, $T_1$ and $T_2$, as functions of external flux is necessary in order to produce a good phenomenology. Such an approach may be used to parameterise the master equation framework presented in this work and to assess its effectiveness in modelling decoherence processes on Josephson junction based devices. 


\section{Conclusions} 

The necessity to consider stronger environmental coupling than might be admitted in lowest order Born Approximation, or the effects of a finite bath cut-off frequency, or of a device operating at low temperature, suggest that the standard Born-Markov development of a Master Equation will need to be extended. The most obvious way to do this is through a small parameter expansion, such as the Born series or, as here, by extending the large cut-off limit by developing the model as a series in the small parameter $\omega_0/\Omega$, or similarly by extending a zero temperature limit. We have chosen, here, perhaps the simplest case (that of a finite cut-off), in the certain knowledge that whatever difficulties one finds are very likely to appear in all others such attempts. 

The most obvious consequence of the present anaysis is that the correction obtained by including second order terms (in $\omega_0/\Omega$) in the Master Equation is not an insignificant one, leading to steady-state impurities $1-P(\DO)$ which are twice those predicted by using a first order model. More subtle is the appearance of the external flux $\Phi_x$ entering the Master Equation, not only in the Hamiltonian terms, but also in the second order Lindblad. Indeed with capacitive coupling the external flux is likely to appear in Lindblads at all orders. As the Josephson coupling energy dictates the height of the potential well, and therefore the tunnelling probability, SQUIDs are (notoriously) sensitive to external magnetic fields and so it is reasonable to expect a strong external flux dependence \cite{Everitt2005, Everitt2001, Everitt2001a}; ~\Eq{eq:meqLindbladSecond} shows such a dependence lies also within the dissipator. Although this is largely contrary to the assumptions of quantum control, where it is generally considered to be the case that, for systems with Lindbladian dissipation, control parameters such as $\Phi_x$ will only enter through the Hamiltonian (see, e.g., [\onlinecite{Rooney2016}]), it is evident from the form of Eq. 21 that $\Phi_x$ can play an important role in dissipation. That the dissipator will in general a function of control parameters has been pointed out previously \cite{D'Alessandro2015}, the current analysis shows they may not all enter at the same order. Furthermore, we have shown that the second order correction to the master equation has a surprisingly large effect. Hence, an understanding of this phenomenon and the role of $\Phi_x$ will be of importance to those working on Josephson junction based devices especially for emerging quantum technologies.
 
 	Recent analysis of the Quantum Brownian Motion (QBM) system indicates that both regular and anomalous diffusion parameters show a logarithmic divergence on bath cut-off frequency $\Omega$, implying a finite cut-off. It thus makes sense to consider a series solution, to different orders of $\omega_0/ \Omega$, if only to check that the common first order truncation is accurate. It is not surprising that, as with QBM, it is necessary to add extra terms in order to bring the master equation into Lindblad form and so avoid unphysical system development. However, in our second order approximation, the extra term needed to complete the first order Lindblad $\OpL_1$ is of a lower order than the terms which make up the second order Lindblad $\OpL_2$. This makes the `minimally invasive' argument a difficult one to sustain and so we appear to be left with the choice of abandoning hierarchical checks, reworking a new standard method, or abandoning the Lindblad form for systems such as these. None of which is attractive. 

	With the exception of a quadratic constraining potential, which is simple because the position operator ($\Phi$ here) links only neighbouring states of fixed energy difference $\omega_0$, all other systems are likely to run into the same difficulties we have here. 
\begin{acknowledgments}
MJE would like to thank Kae Nemoto and SNAD would like to thank Todd Tilma for the generous hospitality, valuable discussions, and support whilst visiting them in Tokyo. SNAD and MJE would like to thank Michael Hanks and Jason Ralph for many valuable, and enjoyable, discussions. MJE, KNB and VMD would like to thank the DSTL for their support through the grant \emph{Engineering for Quantum Reliability}.
This paper recognises the use of the `Hydra' High Performance System at Loughborough University.\end{acknowledgments}

\bibliographystyle{apsrev4-1}
\bibliography{Library2}

\end{document}